\documentclass[12pt]{article}

\usepackage{amsfonts}
\usepackage{amssymb}
\usepackage{amsthm}
\usepackage[fleqn]{amsmath}
\usepackage[mathscr]{eucal}

\input{epsf}

\makeatletter
\renewcommand \thesection {\@arabic\c@section}
\renewcommand\thesubsection   {\thesection.\@arabic\c@subsection}
\renewcommand\thesubsubsection{\thesubsection .\@arabic\c@subsubsection}
\renewcommand\theparagraph    {\thesubsubsection.\@arabic\c@paragraph}
\renewcommand\section{\@startsection {section}{1}{\z@}%
                                   {-3.5ex \@plus -1ex \@minus -.2ex}%
                                   {1.9ex \@plus.2ex}%
                                   {\normalfont\large\bfseries\centering}}
\renewcommand\subsection{\@startsection{subsection}{2}{\z@}%
                                     {-2ex\@plus -1ex \@minus -.2ex}%
                                     {1.2ex \@plus .2ex}%
                                    {\normalfont\normalsize\bfseries\centering}
}
\renewcommand\subsubsection{\@startsection{subsubsection}{3}{\z@}%
                                     {-2ex\@plus -1ex \@minus -.2ex}%
                                     {.5ex \@plus .2ex}%
                                     {\normalfont\normalsize\em}}
\renewcommand\paragraph{\@startsection{paragraph}{4}{\z@}%
                                    {3.25ex \@plus1ex \@minus.2ex}%
                                    {-1em}%
                                    {\normalfont\normalsize\em}}
\renewcommand\subparagraph{\@startsection{subparagraph}{5}{\parindent}%
                                       {3.25ex \@plus1ex \@minus .2ex}%
                                       {-1em}%
                                      {\normalfont\normalsize\em}}
\makeatother

\newcounter{subequation}
	\newenvironment{subequation}%
	{\addtocounter{equation}{-1}%
	\stepcounter{subequation}%
	\begin{equation}}%
	{\end{equation}%
}

\newcommand{\beq}{\begin{equation}}
\newcommand{\eeq}{\end{equation}}
\newcommand{\bseq}{\begin{subequation}}
\newcommand{\eseq}{\end{subequation}}
\newcommand{\bea}{\begin{eqnarray}}
\newcommand{\eea}{\end{eqnarray}}

\newcommand{\cN}{{\cal N}}

\newcommand{\Cset}{{\mathbb C}}

\newcommand{\Nset}{{\mathbb N}}

\newcommand{\Rset}{{\mathbb R}}

\newcommand{\dd}{{\mathrm{d}}}

\newcommand{\Arm}{{\mathrm{A}}}
\newcommand{\Brm}{{\mathrm{B}}}
\newcommand{\Crm}{{\mathrm{C}}}
\newcommand{\Srm}{{\mathrm{S}}}
\newcommand{\BSrm}{{\mathrm{BS}}}
\newcommand{\BIrm}{{\mathrm{BI}}}

\begin{document}

\title{Convergent perturbative power series solution of the stationary 
       Maxwell--Born--Infeld field equations with regular sources}

\author{M.K.-H. KIESSLING\\
\textit{Department of Mathematics,}\\
\textit{Rutgers, The State University of New Jersey,}\\
\textit{110 Frelinghuysen Rd., Piscataway, NJ 08854}}

\date{}

\maketitle

\begin{abstract}
 The stationary Maxwell-Born-Infeld field equations of electromagnetism with regular
sources $\rho\in (C^\alpha_0\cap L^1)(\Rset^3)$ and 
$j\in (C^\alpha_0\cap L^1)(\Rset^3)$ (componentwise) are solved using
a perturbation series expansion in powers of Born's electromagnetic constant.
	The convergence in $C^{1,\alpha}_0$ of the power series for the fields is proved 
with the help of Banach algebra arguments and complex analysis.
 The finite radius of convergence depends on the ``$C^{1,\alpha}_0$ size'' of both, the Coulomb
field generated by $\rho$ and the Amp\`ere field generated by $j$.
 No symmetry is assumed.
\end{abstract}

\bigskip
\bigskip
\bigskip
\centerline{Version of Sept. 09, 2010; typos corrected on January 16, 2011. }
\centerline{To appear in J. Math. Phys. (2011)} 
\vfill
\hrule
\smallskip
\noindent
\copyright{2011} The copyright for this preprint resides with the author. 
Its reproduction, in its entirety, for non-commercial purposes is permitted.
The copyright for the published version resides with AIP. 
\newpage
%

\section{Introduction}

 In a previous paper \cite{CarKieCPAM}, the author and Holly Carley developed a
constructive approach to the prescribed mean-curvature equation \cite{GilbargTrudinger}
\begin{equation}
\pm\nabla\cdot\frac{\nabla u}{\sqrt{1\pm|\nabla u|^2}} = nH
\label{eq:Hpde}
\end{equation}
for hypersurfaces which are graphs of some scalar function $u$ over $\Rset^n$, having mean-curvature 
$H\in (C^\alpha_0\cap L^1)(\Rset^3)$ suitably small. 
 In (\ref{eq:Hpde}) the $+$ sign refers to hypersurfaces in Euclidean $\Rset^{n+1}$, while the $-$ 
sign refers to spacelike hypersurfaces in Minkowski spacetime $\Rset^{1,n}$.
 This latter setting, when $n=3$, is equivalent to the electrostatic Maxwell--Born--Infeld problem
for the electric potential $\phi\propto u$ with regular charge density $\rho\propto H$ and
small Born constant $\beta$.
 In the present paper this constructive approach is generalized to the stationary 
Maxwell--Born--Infeld 
field equations for the electric and magnetic potentials $\phi$ and $A$, or rather their electric 
and magnetic fields $E$ and $B$, with regular charge density $\rho$ and current vector-density $j$.
 We here are interested only in electromagnetic fields in unbounded space $\Rset^3$ with vanishing 
conditions at spatial infinity.

 Since the Maxwell--Born--Infeld field equations were proposed as a nonlinear remedy for the 
infinite field energies and momenta incurred with point charge sources in the linear 
Maxwell--Maxwell field equations (the Maxwell--Lorentz theory), our study of the 
Maxwell--Born--Infeld field equations with regular source terms $\rho$ and $j$  
may seem irrelevant from a fundamental physical perspective.
 However, sufficiently far away from delta function sources the fields should be practically
indistinguishable from regularizations thereof, yet very little is known about solving the 
Maxwell--Born--Infeld field equations for non-special choices of source terms. 
 
 In the next section we briefly summarize the stationary Maxwell--Maxwell and 
Maxwell--Born--Infeld field equations in dimensionless units in which the speed of light $c=1$.
 There we also explain our terminology.
 Section 3 puts forward our perturbative strategy to solve these equations. 
 In section 4 we prove the convergence of the perturbative, formal power series solution. 
 In section 5 we specialize to the electrostatic and magnetostatic subproblems.
 Section 6 concludes the main part of this paper with a brief summary and an outlook on applications.
 The appendix summarizes some important details concerning the recurrence relation for the 
coefficients of the perturbative power series which were supplied by my colleague Doron Zeilberger
in collaboration with Shalosh B. Ekhad III.

\newpage
\section{Stationary electromagnetic Maxwell--Born--Infeld theory}

 Maxwell's classical electromagnetic theory comprises Coulomb's law, Faraday's law,
the Amp\`ere--Maxwell law, and the unnamed law of the absence of magnetic charges.
 In stationary situations the Amp\`ere--Maxwell law reduces to Amp\`ere's law, and
Faraday's law to its stationary special case.
 Coulomb's law states that an electrostatic charge density $\rho$ in $\Rset^3$ is the 
source for the electric displacement field $D$,
\begin{equation}
        {\nabla}\cdot D
=
         4 \pi \rho
\phantom{blablablabl}\mathrm{Coulomb's\ law}
\,,
\label{eq:COULOMBlaw}
\end{equation}
while Faraday's law says that the electric field $E$, when stationary, is curl-free:
\begin{equation}
\nabla\times{E} = {0} 
\phantom{blablablabla} \mathrm{Faraday's\ law\ (stationary)}
\,.\label{eq:FARADAYlawstatic}
\end{equation}
 Similarly, Amp\`ere's law states that a solenoidal electric current vector-density $j$ in 
$\Rset^3$ is the source for the magnetic excitation field $H$,
\begin{equation}
        {\nabla}\times H
=
         4 \pi j
\phantom{blablabla}\mathrm{Amp\grave{e}re's\ law}
\,,
\label{eq:AMPERElaw}
\end{equation}
while the law of the absence of sources of the magnetic induction field $B$ reads
\begin{equation}
        {\nabla}\cdot B
=
      0
\phantom{blablablablab}\mathrm{no-magnetic-charges\ law}
\,.
\label{eq:divBlaw}
\end{equation}

  The four fields $E,D;B,H$ need to be linked by an ``aether law,'' with the help
of which one can eliminate one of the electric-, and one of the magnetic-type fields
from the equations and obtain a closed, nonlinear set of first-order pde for the two 
remaining fields.
 Note that for us ``aether'' is just convenient shorthand for ``electromagnetic vacuum;'' 
of course, for Maxwell and his contemporaries it had a substantial meaning until it was 
demolished by Einstein.
 Maxwell's law of the ``pure aether'' reads
\begin{equation}
E 
= 
D
\phantom{blablabla}\mathrm{Maxwell's\ E-law};
\phantom{blablablablab}
\label{eq:MAXWELLlawEofD}
\end{equation}
\begin{equation}
B 
= 
H
\phantom{blablabla}\mathrm{Maxwell's\ B-law};
\phantom{blablablablab}
\label{eq:MAXWELLlawBofH}
\end{equation}
whereas the aether law of Born and Infeld \cite{BornInfeldB} says
\begin{equation}
E 
= 
\frac{D - \beta^4 D\cdot H H}{ \sqrt{ 1 + \beta^4 (|D|^2 -|H|^2 ) - \beta^8(H\cdot D)^2 }} 
\phantom{blabla}\mathrm{Born-Infeld's\ E-law};
\label{eq:BORNINFELDlawEofDandH}
\end{equation}
\begin{equation}
B 
= 
\frac{H + \beta^4 H\cdot D D}{ \sqrt{ 1 + \beta^4 (|D|^2 -|H|^2 ) - \beta^8(H\cdot D)^2 }} 
\phantom{blabla}\mathrm{Born-Infeld's\ B-law}.
\label{eq:BORNINFELDlawBofDandH}
\end{equation}
\newpage

\noindent
 In (\ref{eq:BORNINFELDlawEofDandH}) and (\ref{eq:BORNINFELDlawBofDandH}), 
$\beta\in (0,\infty)$ is a hypothetical new constant of nature
(in the dimensionless notation of \cite{KieJSPa}), which we have called Born's aether constant.
 In the limit $\beta\to 0$ the Born--Infeld law goes over into Maxwell's aether law.

 To avoid any ambiguity as to which set of field equations we refer to in this paper (and elsewhere)
we will always use the combination ``Maxwell--whose aether law.'' 
 Thus, in particular, we will speak of the Maxwell--Maxwell equations when normally one just refers
to the Maxwell equations.

\section{Formal perturbation theory}

 Assume that the sources $\rho\in (C^\alpha_0\cap L^1)(\Rset^3)$, and 
$j\in (C^\alpha_0\cap L^1)(\Rset^3)$ componentwise. 
 We seek stationary solutions in unbounded space $\Rset^3$ of the
electromagnetic Maxwell--Born--Infeld field equations satisfying vanishing 
conditions at spatial infinity.

 We begin by recalling the just mentioned fact that for $\beta\to 0$ the system of
Maxwell--Born--Infeld field equations reduces to the system of Maxwell--Maxwell field equations, 
which for the stipulated class of sources and asymptotic conditions
are uniquely solved by $E_\Crm = D_\Crm$ and $B_{\Arm} = H_{\Arm}$, where
\begin{equation}
\label{eq:CoulombD}
D_\Crm(x):= - \nabla \int\frac{\rho(y)}{|x-y|}\,d^3 y
\end{equation}
\begin{equation}
\label{eq:AmpereH}
H_{\Arm}(x):=  \nabla\times \int\frac{j(y)}{|x-y|}\,d^3 y.
\end{equation}
Here, the subscripts ${}_\Crm$ and ${}_{\Arm}$ stand for Coulomb and Amp\`ere 
respectively.
 This suggests that for $\beta>0$ we may seek a solution  to the Maxwell--Born--Infeld field 
equations for the same sources and asymptotic conditions as for the Maxwell--Maxwell field 
equations by setting $D_{\BIrm} = D_\Crm + D_\beta$ and $H_{\BIrm} = H_{\Arm}+H_\beta$, with 
$\lim_{\beta\to 0}D_\beta = 0$ and $\lim_{\beta\to 0}H_\beta = 0$, 
from which the electric field $E_{\BIrm}$ and magnetic induction field $B_{\BIrm}$ are obtained 
through the Born--Infeld aether law.

 To determine $D_\beta$ and $H_\beta$ we first note that, since ${\nabla}\cdot D_\Crm = 4 \pi \rho$,
we have that
\begin{equation}
\nabla \cdot D_\beta=0,
\label{eq:divDb}
\end{equation}
and since ${\nabla}\times H_{\Arm} = 4 \pi j$, we have that 
\begin{equation}
\nabla \times H_\beta=0.
\label{eq:curlHb}
\end{equation}
 Thus, $D_\beta$ is a curl field, while $H_\beta$ is a gradient field. 
\newpage

 Incidentally, since moreover $D_\Crm$ is a gradient field and $H_\Arm$ a curl field, we now see that 
the Ansatz $D_{\BIrm} = D_\Crm + D_\beta$ and $H_{\BIrm} = H_{\Arm}+H_\beta$ corresponds precisely to the
Helmholtz decomposition for the fields $D_{\BIrm}$ and $H_{\BIrm}$.
 For each field we have one of the two Helmholtz components and seek the remaining one.
 
 Coming back to the determination of $D_\beta$ and $H_\beta$, we now need a curl equation for 
$D_\beta$ and a divergence equation for $H_\beta$.
 So, using (\ref{eq:BORNINFELDlawEofDandH}) to eliminate $E$ in favor of $D$ and $H$ in 
(\ref{eq:FARADAYlawstatic}), and using (\ref{eq:BORNINFELDlawBofDandH}) to eliminate $B$ 
in favor of $D$ and $H$ in (\ref{eq:divBlaw}), 
we obtain a system of nonlinear vector pde for $D_\beta$ and $H_\beta$ which is closed
\emph{conditioned on $D_\Crm$ and $H_{\Arm}$ being given}. 
 This systems of conditional vector pde reads
\begin{equation}
   \label{curlEofDandHeqn}
   \nabla\times 
   \frac{D - \beta^4 D\cdot H H}{ \sqrt{ 1 + \beta^4 (|D|^2 -|H|^2 ) - \beta^8(H\cdot D)^2 }} 
   =0,
\end{equation}
\begin{equation}
   \label{divBofDandHeqn}
   \nabla\cdot
   \frac{H + \beta^4 H\cdot D D}{ \sqrt{ 1 + \beta^4 (|D|^2 -|H|^2 ) - \beta^8(H\cdot D)^2 }} 
   =0,
\end{equation}
where $D$ and $H$ stand for $D_\Crm + D_\beta$ and $H$ for $H_{\Arm}+H_\beta$, respectively.

 The analyticity properties of the nonlinearities now suggest in particular that 
a solution pair $D_\beta,H_\beta$ for the nonlinear first-order vector 
problem for small but finite $\beta$ is itself analytic in $\beta$, i.e.
satisfies a power series Ansatz, more specifically
\begin{equation}
   \label{eq:powerSERIESforDb}
   D_\beta(x) = \sum_{k=1}^\infty \beta^{4k} D^{(k)}(x),
\end{equation}
\begin{equation}
   \label{eq:powerSERIESforHb}
   H_\beta(x) = \sum_{k=1}^\infty \beta^{4k} H^{(k)}(x),
\end{equation}
with each $D^{(k)}$ and $H^{(k)}$ independent of $\beta$.
 Inserting this power series Ansatz into the pair of equations 
(\ref{curlEofDandHeqn}), (\ref{divBofDandHeqn}) and sorting according to powers of 
$\beta$, we find a hierarchy of linear equations, the $k$-th members of which read:
\begin{equation}
   \label{eq:CURLlawDk}
   \nabla\times  D^{(k)} = \nabla\times \sum_{h=1}^k \Big(X^{(h)}D^{(k-h)} + Y^{(h)}H^{(k-h)}\Big);
   \quad k\in\Nset
\end{equation}
(together with $\nabla\cdot D^{(k)}=0$), and 
\begin{equation}
   \label{eq:DIVlawHk}
   \nabla\cdot H^{(k)} = \nabla\cdot \sum_{h=1}^k \Big(X^{(h)}H^{(k-h)} - Y^{(h)}D^{(k-h)}\Big);
   \quad k\in\Nset
\end{equation}
(together with $\nabla\times H^{(k)}=0$); here, $D^{(0)}:=D_\Crm$ and $H^{(0)}:=H_\Arm$,  and 
$X^{(h)}$ and $Y^{(h)}$ 
\newpage

\noindent
are the coefficients of $\beta^{4h}$ in the power series expansions of the two formal functions
$$
 \beta^4\mapsto  - 1/\sqrt{1 + \beta^4 (|D|^2 -|H|^2 ) - \beta^8(H\cdot D)^2}
$$ 
and
$$
 \beta^4\mapsto\beta^4 D\cdot H/\sqrt{1+\beta^4 (|D|^2 -|H|^2 ) - \beta^8(H\cdot D)^2},
$$
respectively, with $D=D_\Crm+D_\beta$ and $H= H_\Arm+H_\beta$.
 Explicitly, and including $h=0$, we have $X^{(0)}=-1$, while for $h\in\Nset$ we have
\begin{equation}
   \label{eq:XhDEF}
   X^{(h)} = 
   - \sum_{j=1}^h M_j\!\sum_{l=0}^j(-1)^{j-l}\left(\begin{matrix}j\\ l\end{matrix}\right)
     Z_{h,j,l},
\end{equation}
with
\begin{equation}
   \label{eq:ZhjnullDEF}
   Z_{h,j,0}=
   \sum_{|b|_{4j} = h -2j} 
\prod_{n=1}^{2j}D^{(b_{2n-1})}{\cdot}H^{(b_{2n})},
\end{equation}
\begin{equation}
   \label{eq:ZhjjDEF}
   Z_{h,j,j}=
   \sum_{|a|_{2j} = h -j} 
\prod_{m=1}^j(D-H)^{(a_{2m-1})}{\cdot}(D+H)^{(a_{2m})},
\end{equation}
and, for $0<l<j$,
\begin{equation}
   \label{eq:ZhjlDEF}
   Z_{h,j,l}=
   \sum_{|a|_{2l}+|b|_{4(j-l)} \atop = h +l-2j} 
\prod_{m=1}^l(D-H)^{(a_{2m-1})}{\cdot}(D+H)^{(a_{2m})}
\prod_{n=1}^{2(j-l)}D^{(b_{2n-1})}{\cdot}H^{(b_{2n})}.
\end{equation}
 Here, $|\ell|_{K} := \sum\limits_{i=1}^{K} \ell_i$, with 
$\ell_i$ taking any non-negative integer values; so in particular, $\sum_{|\ell|_K<0}...=0$.
 Also, $M_0:=1$ and
\begin{equation}
   \label{eq:Mpm}
   M_j = (-1)^j\frac{(2j-1)!!}{j!2^j}; \quad j\in\Nset
\end{equation}
are the Maclaurin coefficients for $1/\sqrt{1+ z}$.
 Having $X^{(h)}$ for $h=0,1,2,...$, we find
\begin{equation}
   \label{eq:YhDEF}
   Y^{(h)} = - \sum_{g=0}^{h-1}  X^{(h-1-g)} \sum_{|a|_2=g} D^{(a_1)}{\cdot}H^{(a_2)};
\quad
h\in\Nset.
\end{equation}
 Note that $X^{(h)}$ and $Y^{(h)}$ contain only terms of order $<h$ of 
$|D_\beta|^2$, $|H_\beta|^2$, and $D_\beta\cdot H_\beta$. 
\newpage

 The expressions for $X^{(k)}$ and $Y^{(k)}$ look somewhat unwieldy, but since there is
scant empirical evidence (if any!) for the need of corrections to the solutions of the 
Maxwell--Maxwell field equations in the regime of classical phenomena, for many practical
applications it may suffice to just compute the first correction term to the Coulomb and 
Amp\`ere fields.
 These are of a more manageable structure.
 In particular, $D^{(1)}$ satisfies
\begin{equation}
   \label{eq:CURLlawDone}
   \nabla\times  D^{(1)}
=    \nabla\times \Big( \tfrac12 \big(|D_\Crm|^2-|H_\Arm|^2\big)D_\Crm + D_\Crm\cdot H_\Arm H_\Arm \Big) 
\,,
\end{equation}
together with $\nabla\cdot D^{(1)}=0$, and $H^{(1)}$ satisfies
\begin{equation}
   \label{eq:DIVlawHone}
   \nabla\cdot H^{(1)}
=
   \nabla\cdot  \Big(  \tfrac12 \big(|D_\Crm|^2-|H_\Arm|^2\big)H_\Arm - D_\Crm\cdot H_\Arm D_\Crm \Big) 
\,,
\end{equation}
together with $\nabla\times H^{(1)}=0$.

 As to the solvability of the linear first-order PDE (\ref{eq:CURLlawDk}), (\ref{eq:DIVlawHk}),
assuming that their right-hand sides are in $C^{\alpha}_0$, this pair of PDE has
a unique solution in $C^{1,\alpha}_0$ given by
\begin{equation}
   \label{eq:DkSOL}
   D^{(k)} =
   {\mathbf P} \sum_{h=1}^k \Big(X^{(h)}D^{(k-h)} + Y^{(h)}H^{(k-h)}\Big),\qquad k\in\Nset
\end{equation}
\begin{equation}
   \label{eq:HkSOL}
   H^{(k)} =
   {\mathbf Q} \sum_{h=1}^k \Big(X^{(h)}H^{(k-h)} - Y^{(h)}D^{(k-h)}\Big),\qquad k\in\Nset
\end{equation}
where $\mathbf{P}\!: C^{1,\alpha}_{0}\!\to C^{1,\alpha}_{0}$ projects onto the solenoidal 
subspace of $C^{1,\alpha}_{0}$, and where $\mathbf{Q}\!: C^{1,\alpha}_{0}\!\to C^{1,\alpha}_{0}$ 
projects onto the curl-free subspace of $C^{1,\alpha}_{0}$.
 More explicitly, for a vector field $V^{(k)}\in C^{1,\alpha}_{0}$ with
$\nabla\cdot V^{(k)}\in C^{0,\alpha}_{0}\cap L^{1}$, we have
\begin{equation}
   \label{eq:DsolEXPLk}
	 {\mathbf P}V^{(k)}(x) =
	 V^{(k)}(x) 
	 + \nabla \int \frac{{\textstyle{\frac{1}{4\pi}}}\nabla\cdot V^{(k)}(y)}{|x-y|}\,\dd^3 y,
\end{equation}
and when $\nabla\times V^{(k)}\in C^{0,\alpha}_{0}\cap L^{1}$, we have
\begin{equation}
   \label{eq:HsolEXPLk}
	 {\mathbf Q}V^{(k)}(x) =
	 V^{(k)}(x) 
    - \nabla \times\int \frac{{\textstyle{\frac{1}{4\pi}}}\nabla\times V^{(k)}(y)}{|x-y|}\,\dd^3 y.
\end{equation}

 To summarize, so far we have seen that the Ansatz
$D= D_\Crm+ \sum_{k=1}^\infty \beta^{4k} D^{(k)}$ 
and
$H= H_\Arm+ \sum_{k=1}^\infty \beta^{4k} H^{(k)}$,
with $D^{(k)}$ and $H^{(k)}$ for $k\in\Nset$ recursively given by 
(\ref{eq:DkSOL}) and (\ref{eq:HkSOL}), yields a 
\emph{formal} series solution pair for the stationary Maxwell--Born--Infeld field equations
with regular sources.
 We next prove its convergence in $C^{1,\alpha}_{0}$ for small enough $\beta$, given the
sources $\rho\in C^{\alpha}_{0}$ and $j\in C^{\alpha}_{0}$.

\section{Rigorous perturbation theory $\scriptscriptstyle{\mathrm{modulo\ some\ MAPLE\ calculations}}$}

\subsection{$C^{1,\alpha}_0$ consistency of the formal perturbation series solution}

 We first confirm that the formal series 
$D_\beta = \sum_{k=1}^\infty \beta^{4k} D^{(k)}$ and
$H_\beta = \sum_{k=1}^\infty \beta^{4k} H^{(k)}$ are $C^{1,\alpha}_{0}$ series.
 For this we have to check that all $D^{(k)}$ and $H^{(k)}$ as given in 
(\ref{eq:DsolEXPLk}) and (\ref{eq:HsolEXPLk}) are in $C^{1,\alpha}_{0}$.
 But this follows inductively from the facts 
that $D^{(0)}=D_\Crm\in C^{1,\alpha}_{0}$ and $H^{(0)}=H_\Arm\in C^{1,\alpha}_{0}$, 
that $C^{1,\alpha}_{0}$ is a Banach algebra, and that 
$\mathbf{P}:C^{1,\alpha}_{0}\to C^{1,\alpha}_{0}$ and 
$\mathbf{Q}:C^{1,\alpha}_{0}\to C^{1,\alpha}_{0}$; hence,
each partial sum of 
$D_\beta = \sum_{k=1}^\infty \beta^{4k} D^{(k)}$ is in $C^{1,\alpha}_{0}$ and so is
each partial sum of 
$H_\beta = \sum_{k=1}^\infty \beta^{4k} H^{(k)}$.

\subsection{Absolute convergence of the formal perturbation series solution}

 Since
$\|D_\beta\|\leq \sum_{k=1}^\infty \beta^{4k} \|D^{(k)}\|$ 
and
$\|H_\beta\|\leq \sum_{k=1}^\infty \beta^{4k} \|H^{(k)}\|$
in the sense of partial sums, to prove absolute convergence of the 
power series for $D_\beta$ and $H_\beta$ it suffices to show that 
$\sum_{k=1}^\infty \beta^{4k} \big(\|D^{(k)}\|+\|H^{(k)}\|\big)<\infty$ for sufficiently 
small $\beta$, given $\rho$ and $j$.

 We first estimate $\|D^{(k)}\| + \|H^{(k)}\|$ for $k\in\Nset$ 
in terms of the $\|D^{(l)}\| + \|H^{(l)}\|$ for all $l<k$. 
 Beginning with $\|D^{(k)}\|$, for $k\in\Nset$ we find

\vskip-.5truecm
\begin{eqnarray}
   \label{eq:DkNORMid}
\|D^{(k)}\| &=&
   \|{\mathbf P} {\sum}_{h=1}^k \Big(X^{(h)}D^{(k-h)} + Y^{(h)}H^{(k-h)}\Big)\|\\
\label{DkNORMestA}
&\leq& 
\| {\sum}_{h=1}^k \Big(X^{(h)}D^{(k-h)} + Y^{(h)}H^{(k-h)}\Big)\|\\
\label{DkNORMestB}
&\leq&
{\sum}_{h=1}^k \| \Big(X^{(h)}D^{(k-h)} + Y^{(h)}H^{(k-h)}\Big)\|\\
\label{DkNORMestC}
&\leq&
{\sum}_{h=1}^k \Big(\| X^{(h)}D^{(k-h)}\| + \|Y^{(h)}H^{(k-h)}\|\Big)\\
\label{DkNORMestD}
&\leq&
{\sum}_{h=1}^k \Big(\| X^{(h)}\|\|D^{(k-h)}\| + \|Y^{(h)}\|\|H^{(k-h)}\|\Big).
\end{eqnarray}
 Here, inequality (\ref{DkNORMestA}) holds because 
$\mathbf{P}:C^{1,\alpha}_{0}\to C^{1,\alpha}_{0}$ is a projector, 
inequalities (\ref{DkNORMestB}) and (\ref{DkNORMestC}) are just the triangle inequality,
while (\ref{DkNORMestD}) is valid in Banach algebras (here $C^{1,\alpha}_{0}$).
 Similarly, for $\|H^{(k)}\|$ and all $k\in\Nset$ we find
\begin{eqnarray}
   \label{eq:HkNORMest}
\|H^{(k)}\| \leq 
{\sum}_{h=1}^k \Big(\| X^{(h)}\|\|H^{(k-h)}\| + \|Y^{(h)}\|\|D^{(k-h)}\|\Big).
\end{eqnarray}
 Adding these estimates for $\|D^{(k)}\|$ and $\|H^{(k)}\|$ yields 
\begin{eqnarray}
   \label{eq:DkHkNORMest}
\|D^{(k)}\| + \|H^{(k)}\| 
\leq 
{\sum}_{h=1}^k \Big(\| X^{(h)}\|+ \|Y^{(h)}\|\Big)\Big(\|D^{(k-h)}\|+\|H^{(k-h)}\| \Big).
\end{eqnarray}
 Since $\|X^{(h)}\|$ and $\|Y^{(h)}\|$  depend in turn on $\|D^{(l)}\|$ and $\|H^{(l)}\|$, we need 
to carry on and estimate $\|X^{(h)}\| +\|Y^{(h)}\|$ in terms of the $\|D^{(l)}\| + \|H^{(l)}\|$ 
for all $l<k$. 
 As to $\|X^{(h)}\|$, we have $\|X^{(0)}\|=1$ and
\begin{equation}
   \label{eq:XhNORMest}
\|X^{(h)}\| 
\leq 
    \sum_{j=1}^h |M_j|\!\sum_{l=0}^j \left(\begin{matrix}j\\ l\end{matrix}\right) \|Z_{h,j,l}\|
\end{equation}
and, for $h\in \Nset$,
\begin{eqnarray}
   \label{eq:ZhjnullNORMestA}
\|Z_{h,j,0}\|
\!\!&\leq&\!\!
   \sum_{|b|_{4j} = h -2j} 
\prod_{n=1}^{2j}\|D^{(b_{2n-1})}\|\|H^{(b_{2n})}\| \\ 
   \label{eq:ZhjnullNORMestB}
\!\!&\leq&\!\!
   \sum_{|b|_{4j} = h -2j} 
\prod_{n=1}^{4j}\Big(\|D^{(b_{n})}\|+\|H^{(b_{n})}\| \Big),
\\ 
   \label{eq:ZhjjNORMest}
\|Z_{h,j,j}\|
\!\!&\leq&\!\!
   \sum_{|a|_{2j} = h -j} 
\prod_{m=1}^j\|(D-H)^{(a_{2m-1})}\|\|(D+H)^{(a_{2m})}\|\\ 
\!\!&\leq&\!\!
   \sum_{|a|_{2j} = h -j} 
\prod_{m=1}^{2j}\big(\|D^{(a_m)}\|+\|H^{(a_{m})}\|\big),
\end{eqnarray}
and if $j>1$, which can happen only when $h\geq 2$, then for $0<l<j$ we have
\begin{eqnarray}
   \label{eq:ZhjlNORMest}
\hskip-.7truecm
\| Z_{h,j,l}\|
\!\!&\leq&\!\!\!\!\!\!
   \sum_{|a|_{2l}+|b|_{4(j-l)} \atop = h +l-2j} 
\prod_{m=1}^l\!\|(D-H)^{(a_{2m-1})}\|\|(D+H)^{(a_{2m})}\|
\!\!\prod_{n=1}^{2(j-l)}\!\|D^{(b_{2n-1})}\|\|H^{(b_{2n})}\|\\
\!\!&\leq&\!\!\!\!
   \sum_{|a|_{2l}+|b|_{4(j-l)} \atop = h +l-2j} 
\prod_{m=1}^{2l}\big(\|D^{(a_{m})}\|+\|H^{(a_{m})}\|\big)
\prod_{n=1}^{4(j-l)}\Big(\|D^{(b_{n})}\|+\|H^{(b_{n})}\| \Big).
\end{eqnarray}
 Finally, as to $\|Y^{(h)}\|$, we have
\begin{eqnarray}
   \label{eq:YhNORMestA}
\|Y^{(h)}\| 
\!\!&\leq &\!\!
\sum_{g=0}^{h-1} \| X^{(h-1-g)}\| \sum_{|a|_2=g} \|D^{(a_1)}\|\|H^{(a_2)}\|\\
   \label{eq:YhNORMestB}
\!\!&\leq &\!\!
\sum_{g=0}^{h-1} \| X^{(h-1-g)}\| \sum_{|a|_2=g} \Big(\|D^{(a_1)}\|+\|H^{(a_1)}\|\Big)
\Big(\|D^{(a_2)}\|+\|H^{(a_2)}\|\Big).
\end{eqnarray}

 Having completed our estimate of  $\|D^{(k)}\| + \|H^{(k)}\|$ for $k\in\Nset$ 
in terms of the $\|D^{(l)}\| + \|H^{(l)}\|$ for all $l<k$, we now abbreviate 
$\|D^{(n)}\|+\|H^{(n)}\|= N^{(n)}$ and recast our estimates more user-friendly as
\begin{eqnarray}
   \label{eq:NkNORMest}
N^{(k)}
\leq 
\sum_{h=1}^k \Big(\| X^{(h)}\|+ 
\sum_{g=0}^{h-1} \| X^{(h-1-g)}\| \sum_{|a|_2=g} N^{(a_1)}N^{(a_2)}\Big) N^{(k-h)},
\end{eqnarray}
together with $\|X^{(0)}\|=1$, then
\begin{equation}
   \label{eq:XoneNORMest}
\|X^{(1)}\| 
\leq 
    {\textstyle{\frac12}}
   \sum_{|a|_{2} = 0}\prod_{m=1}^{2}\! N^{(a_{m})}
=     {\textstyle{\frac12}} {N^{(0)}}^2 ,
\end{equation}
and, for $h\geq 2$,
\begin{eqnarray}
   \label{eq:XhNORMestREV}
\|X^{(h)}\| 
&\leq &
    |M_1|\Biggl[ \sum_{|b|_{4} =h -2} \prod_{n=1}^{4}N^{(b_{n})}
+
   \sum_{|a|_{2} = h -1} \prod_{m=1}^{2}\! N^{(a_{m})}
\Biggr]\\
&&
   \nonumber
\!\!\! +  \sum_{j=2}^h |M_j|\Biggl[ \sum_{|b|_{4j}  =h -2j} \prod_{n=1}^{4j}N^{(b_{n})}
+
   \sum_{|a|_{2j} = h -j} \prod_{m=1}^{2j}\! N^{(a_{m})}\Biggr.\\
&&
 \nonumber
\qquad\qquad \Biggl. + \sum_{l=1}^{j-1} \left(\begin{matrix}j\\ l\end{matrix}\right) \!\!
   \sum_{|a|_{2l}+|b|_{4(j-l)} \atop = h +l-2j} \!
\prod_{m=1}^{2l}\!N^{(a_{m})}\!\!\prod_{n=1}^{4(j-l)}\!\! N^{(b_{n})}
\Biggr].
\end{eqnarray}
 By an inductive argument we now estimate each $N^{(k)}$ in terms of 
the $2k+1$-th power of $N^{(0)}=:\cN$.

 In this vein, setting $k=1$ in (\ref{eq:NkNORMest}) we obtain the estimate
\begin{equation}
\label{eq:NoneESTa}
\qquad
N^{(1)} \leq 
\Big(\| X^{(1)}\|+ \cN^2\Big) \cN,
\end{equation}
and for $\| X^{(1)}\|$ we have (\ref{eq:XoneNORMest}), so 
\begin{equation}
\label{eq:NoneESTb}
\qquad
N^{(1)} \leq     {\textstyle{\frac32}} {\cN}^3.
\end{equation}
 Next, suppose that for all $k=1,...,k_*$ there exists some $R_{k}$ such that
\begin{equation}
\label{eq:NnormESTitera}
\qquad
N^{(k)}\leq  R_{k}{\cN}^{2k+1}.
\end{equation}
 Also set $R_0:=1$ (for, $N^{(0)}=\cN$).
 Inserting these estimates into (\ref{eq:NkNORMest}) then yields
\begin{eqnarray}
   \label{eq:NkstarPLUSoneNORMestBB}
N^{(k_*+1)}
\!\leq \!
\sum_{h=1}^{k_*+1} \!\Big[\| X^{(h)}\|+ 
\!\sum_{g=0}^{h-1} \| X^{(h-1-g)}\| {\cN}^{2g+2}\!\!\sum_{|a|_2=g}\! 
R_{a_1}R_{a_2}\Big] R_{k_*+1-h}{\cN}^{2(k_*+1-h)+1}\!\!.
\end{eqnarray}
 As to the $\|X^{(h)}\|$, for $h=1$ we already have the estimate (\ref{eq:XoneNORMest}), and
inserting (\ref{eq:NnormESTitera}) into (\ref{eq:XhNORMestREV}) we find for $h\geq 2$ that
\begin{eqnarray}
   \label{eq:XhNORMestREVbbA}
\|X^{(h)}\| 
&\leq &
    |M_1|\Biggl[ \sum_{|b|_{4} =h -2} \prod_{n=1}^{4}R_{b_{n}}\cN^{2b_{n}+1}
+
   \sum_{|a|_{2} = h -1} \prod_{m=1}^{2}\! R_{a_{m}}\cN^{2a_{m}+1}
\Biggr]\\
&&
 \nonumber
\!\!\! + \sum_{j=2}^h |M_j|\Biggl[ \sum_{|b|_{4j}  =h -2j} \prod_{n=1}^{4j}R_{b_{n}}\cN^{2b_{n}+1}
+
   \sum_{|a|_{2j} = h -j} \prod_{m=1}^{2j}\! R_{a_{m}}\cN^{2a_{m}+1}
\Biggr.\\
&&
 \nonumber
\qquad\qquad \Biggl. + \sum_{l=1}^{j-1} \left(\begin{matrix}j\\ l\end{matrix}\right) \!\!
   \sum_{|a|_{2l}+|b|_{4(j-l)} \atop = h +l-2j} \!
\prod_{m=1}^{2l}\!R_{a_{m}}\cN^{2a_m+1}\!\!\prod_{n=1}^{4(j-l)}\!\! R_{b_{n}}\cN^{2b_{n}+1}
\Biggr]\\
\label{eq:XhNORMestREVbbB}
&= &
   \cN^{2h}\Biggl( |M_1|\Biggl[ \sum_{|b|_{4} =h -2} \prod_{n=1}^{4}R_{b_{n}}
+
   \sum_{|a|_{2} = h -1} \prod_{m=1}^{2}\! R_{a_{m}}
\Biggr]\Biggr.\\
&&
   \nonumber
\qquad\qquad 
\Biggl.+ \sum_{j=2}^h |M_j|\Biggl[ \sum_{|b|_{4j}  =h -2j} \prod_{n=1}^{4j}R_{b_{n}}
+
   \sum_{|a|_{2j} = h -j} \prod_{m=1}^{2j}\! R_{a_{m}}
\Biggr.\\
&&
   \nonumber
\qquad\qquad \qquad\qquad 
\Biggl. + \sum_{l=1}^{j-1} \left(\begin{matrix}j\\ l\end{matrix}\right) \!\!
   \sum_{|a|_{2l}+|b|_{4(j-l)} \atop = h +l-2j} \!
\prod_{m=1}^{2l}\!R_{a_{m}} \!\!\prod_{n=1}^{4(j-l)}\!\! R_{b_{n}}
\Biggr]\Biggr).
\end{eqnarray}
 Estimates (\ref{eq:XoneNORMest}) and
(\ref{eq:XhNORMestREVbbA}), (\ref{eq:XhNORMestREVbbB}) state that
$\forall h\in\Nset: \exists C_h>0$ such that $\|X^{(h)}\| \leq C_h\cN^{2h}\!$.
 This also holds for $h=0$, with $C_0=1$.
 Inserted back into (\ref{eq:NkstarPLUSoneNORMestBB}) this shows
now that (\ref{eq:NnormESTitera}) is true also for $k=k_*+1$, and since 
$k_*\geq 1$ is arbitrary in this induction step while
(\ref{eq:NoneESTb}) says that (\ref{eq:NnormESTitera}) is true for $k_*=1$, 
it follows that (\ref{eq:NnormESTitera}) is true for all $k\in\Nset$.

 Our inductive argument also yields $R_{k}$ as recursively defined for $k\in\Nset$ by
\begin{eqnarray}
   \label{eq:KkITERA}
R_{k}
= 
\sum_{h=1}^{k} \!\Big(C_{h} + 
\!\sum_{g=0}^{h-1} C_{h-1-g} \!\!\sum_{|a|_2=g}\! 
R_{a_1}R_{a_2}\Big) R_{k-h},
\end{eqnarray}
with $R_0:=1=:C_0$, $C_1 = 1/2$ (\ref{eq:XoneNORMest}), and $C_h$ for $h\geq 2$ given by
coeff($\cN^{2h}$) at rhs(\ref{eq:XhNORMestREVbbB}).
\newpage

 To summarize, so far we have proved that the formal power series solutions for $D_\beta$ 
and $H_\beta$
satisfy
\begin{equation}
\label{eq:powerSERIESestA}
\sum_{k=1}^\infty \beta^{4k} \big(\|D^{(k)}\|+\|H^{(k)}\|\big)
\leq 
\sum_{k=1}^\infty R_k \beta^{4k} \cN^{2k+1}
\end{equation}
with the $R_k$ given as stated in (\ref{eq:KkITERA}) and its ensuing text.
 We next discuss the convergence of rhs(\ref{eq:powerSERIESestA}).

 We begin by noting that the convergence of the series at rhs(\ref{eq:powerSERIESestA}) is
not altered if we multiply it by $\beta^2$ and add a term $\beta^2\cN$ to it. 
 Now defining $\xi:=\beta^2\cN$, as well as $c_1:=1$ and $c_{2k+1}:= R_k$ for $k\in\Nset$,
we see that the convergence of rhs(\ref{eq:powerSERIESestA}) is decided by the convergence 
of the series
\begin{equation}
\label{eq:generatingFCT}
G(\xi) := \sum_{k=0}^\infty c_{2k+1}\xi^{2k+1}
\end{equation}
for $\xi\in\Cset$.
 Note that the formal power series (\ref{eq:generatingFCT}) is just the generating function 
of the $c_{2k+1}$, i.e. $c_{2k+1} = G^{({2k+1})}(0)/({2k+1})!$, formally.
 We need to show that the generating function is analytic about $\xi=0$ with non-zero radius 
of convergence.

 With the help of the recursion relation (\ref{eq:KkITERA}) we readily find that for positive $\xi$
the function $G(\xi)$ is the positive inverse of the locally analytic function $g\mapsto \xi$ given by
\begin{equation}
\label{eq:GimplicitDEF}
\xi = 2g -  \frac{g+g^3}{\sqrt{1-g^2-g^4}},
\end{equation}
which has unit derivative at $g=0$, and so its inverse $\xi\mapsto g$ is analytic 
in an open $\xi$-neighborhood of $\xi=0$, with $G(0)=0$.
 This argument already implies a finite radius of convergence, but 
without any information on its size.
 
 Before we actually determine the radius of convergence we note that
an upper bound for it is readily found by discussing the function 
$g\mapsto \xi$ defined in (\ref{eq:GimplicitDEF}) on its real interval of
definition about zero.
 Clearly, this interval is $(-g_\bullet,g_\bullet)$, where 
$g_\bullet = \sqrt{(\sqrt{5}-1)/2}\approx 0.7861513775$ is 
the positive real zero of the square root term in (\ref{eq:GimplicitDEF}). 
 Furthermore, $g\mapsto \xi$ is odd, concave for positive and convex for negative $g$ values, 
having a unique maximum at about $g^* = 0.4039458281$ and a unique minimum at $-g^*$, yielding
the $\xi$-values $\xi^* = 0.285891853$ for the maximum, and $-\xi^*$ for the minimum; the numerical
values are easily obtained using MAPLE. 
 Clearly, $\xi^*$ is an upper bound to the radius of convergence of the power series for $G(\xi)$.

 We now show that $\xi^*$ is the radius of convergence.
 In fact, the radius of convergence is found amongst
those $\xi$ values closest to $\xi=0$ at which the derivative of 
$g\mapsto \xi$ vanishes (possibly asymptotically should $\xi\to\xi_\infty$ when $g\to\infty$ 
suitably). 
 Now, since any solution $g=G(\xi)$ to (\ref{eq:GimplicitDEF}) with $G(0)=0$ is also a solution 
to the algebraic problem of finding the roots of the polynomial of degree $6$ in $g$ with 
$\xi$-dependent coefficients, given by
\begin{equation}
\label{eq:GimplicitDEFsquare}
P(g|\xi) = (\xi - 2g)^2({1-g^2-g^4}) -(g+g^3)^2,
\end{equation}
the asymptotic scenario does not occur and all one needs to do is to locate the finitely many
zeros of the $g$-derivative of $g\mapsto \xi$, which reads 
\begin{equation}
\label{eq:xiOFgDERIVATIVE}
\frac{d\xi}{dg} = 
-\frac{1+ 3g^2-g^4-g^6-2\sqrt{1-g^2-g^4}^{\,3}}{\sqrt{1-g^2-g^4}^{\,3}}.
\end{equation}
 Clearly, a zero of r.h.s.(\ref{eq:xiOFgDERIVATIVE}) 
is also a zero of the order-$12$ polynomial $\widetilde{P}(g)$
obtained by multiplying the numerator of r.h.s.(\ref{eq:xiOFgDERIVATIVE}) 
by $1+ 3g^2-g^4-g^6 +2\sqrt{1-g^2-g^4}^3$,~viz.
\begin{equation}
\label{eq:xiOFgDERIVATIVEnumSQUARE}
\widetilde{P}(g) = 
(1+ 3g^2-g^4-g^6)^2-4(1-g^2-g^4)^3.
\end{equation}
 By the fundamental theorem of algebra, $\widetilde{P}(g)$ has twelve zeros, counted in multiplicity; 
six of these are zeros of ${d\xi}/{dg}$ given in (\ref{eq:xiOFgDERIVATIVE}).
 With the help of MAPLE one finds that if $g_0$ is a zero of ${d\xi}/{dg}$ given in 
(\ref{eq:xiOFgDERIVATIVE}), then 
\begin{equation}
\label{eq:zerosOFxiOFgDERIVATIVE}
g_0\in\pm\{ 0.4039458281,\, 0.07758059914\pm i1.387412147 \}.
\end{equation}
 Writing $G^{-1}$ for r.h.s.(\ref{eq:GimplicitDEF}), we now have 
\begin{equation}
\label{eq:ABSxiVALUESatTHEzeros}
|G^{-1}(g_0)|\in\{0.285891853,\, 3.235626655\},
\end{equation}
and the smaller of these two values is the radius of convergence of $G(\xi)$. 

 Lastly, we translate the radius of convergence $\xi^*= 0.285891853$ of the series
$G(\xi)$ given in (\ref{eq:generatingFCT}) into a sufficient criterion of convergence
of our formal perturbation series 
$D= D_\Crm+ \sum_{k=1}^\infty \beta^{4k} D^{(k)}$ 
and
$H= H_\Arm+ \sum_{k=1}^\infty \beta^{4k} H^{(k)}$.
 Namely, if 
\begin{equation}
\label{eq:betaBOUND}
\beta^2 (\|D_\Crm\| + \|H_\Arm\|) < 0.285891853
\end{equation}
then our perturbative power series solution pair for $D$ and $H$ converges to a classical solution 
for the stationary Maxwell--Born--Infeld field equations.

 This completes our convergence proof. 

\newpage

\section{Purely electric or magnetic fields}

 When $j\equiv 0$ or $\rho\equiv 0$, then the perturbation theory for solving the stationary
electromagnetic Maxwell--Born--Infeld equations simplifies considerably.
 In particular, when applying the perturbation theory directly to the electrostatic or 
magnetostatic field equations, better convergence estimates are obtained than by simply
specializing the electromagnetic estimates to these cases.

\subsection{The electrostatic case}

 When $j\equiv 0$ and $H$ vanishes at spatial infinity we have
$B\equiv 0\equiv H$ and the stationary Maxwell--Born--Infeld equations
reduce to the electrostatic Maxwell--Born(--Infeld) equations, comprising
Coulomb's law
\begin{equation}
        {\nabla}\cdot D
=
         4 \pi \rho
\,,
\label{eq:COULOMBlawAGAIN}
\end{equation}
and the stationary Faraday law
\begin{equation}
\nabla\times{E} = {0} 
\,.\label{eq:FARADAYlawstaticAGAIN}
\end{equation}
  The two fields $E$ and $D$ are linked by the aether law of Born \cite{BornA,BornD},
\begin{equation}
E 
= 
\frac{D }{ \sqrt{ 1 + \beta^4 |D|^2 }} 
\phantom{blablabla}\mathrm{Born's\ E-law}.
\label{eq:BORNlawEofD}
\end{equation}
 These equations are covered in \cite{CarKieCPAM}.
 For the convenience of the reader we summarize the main results in the notation of the
present paper.

 Thus, inserting our perturbation theory Ansatz $D= D_\Crm + D_\beta$, with 
$D_\beta=\beta^4D^{(1)}+\beta^8 D^{(2)}+\dots$, into (\ref{eq:BORNlawEofD}) 
and collecting powers of $\beta$, we find that $D^{(k)}$ for $k\in\Nset$ satisfies 
the pair of linear first-order PDE
\begin{eqnarray}
\label{eq:DkDIVlaw}
\nabla\cdot D^{(k)}&=&0
\\
\label{eq:DkCURLlaw}
\nabla\times D^{(k)}& =&
\nabla\times V^{(k)},
\end{eqnarray}
where $V^{(k)}$ is a polynomial in the $D^{(\ell)}$ with $\ell<k$, viz.
\begin{equation}
\label{eq:Vdef}
\qquad
V^{(k)} = 
- \sum_{h=1}^k D^{(k-h)}\sum_{j=1}^h M_j\!\!
\sum_{|\ell|_{2j} =\atop h-j}\prod_{i=1}^jD^{(\ell_{2i-1})}{\cdot}D^{(\ell_{2i})},
\end{equation}
where $|\ell|_{2j}$  and $M_j$ have their earlier assigned meaning.
 The pair of linear first-order PDE (\ref{eq:DkDIVlaw}), (\ref{eq:DkCURLlaw}) has
a unique solution in $C^{1,\alpha}_0$ given by
\begin{equation}
\label{eq:DkSOLnoH}
 D^{(k)} =
{\mathbf P} V^{(k)},\qquad k\in\Nset
\end{equation}
where $\mathbf{P}\!: C^{1,\alpha}_{0}\!\to C^{1,\alpha}_{0}$ projects onto the solenoidal 
subspace of $C^{1,\alpha}_{0}$; see (\ref{eq:DsolEXPLk}).
 The first member of this hierarchy reads explicitly
\begin{equation}
\label{eq:DoneEXPL}
D^{(1)}(x)=
{\textstyle{\frac12}}|D_\Crm|^2D_\Crm 
+\nabla
\int \frac{{\textstyle{\frac{1}{4\pi}}}\nabla\cdot({\textstyle{\frac12}}|D_\Crm|^2D_\Crm)(y)}{|x-y|}d^3y,
\end{equation}
so $\beta^4$(\ref{eq:DoneEXPL}) gives the leading correction to the Coulomb term $D^{(0)}=D_\Crm$.

 In general, the perturbation series $D= \sum_{k=0}^\infty \beta^{4k} D^{(k)}$, 
with $D^{(0)}=D_\Crm$ given by (\ref{eq:CoulombD}) and $D^{(k)}$ for $k\in\Nset$ 
recursively given by (\ref{eq:DkSOLnoH}), is a \emph{formal} series solution for the
electrostatic field problem (\ref{eq:COULOMBlawAGAIN}), (\ref{eq:FARADAYlawstaticAGAIN}) 
with prescribed $\rho$.
 Following essentially the same strategy, in 
\cite{CarKieCPAM} the convergence in $C^{1,\alpha}_{0}$ of the formal power series is proved for 
\begin{equation}
\label{eq:radOFconv}
 \beta^2\|D_\Crm\| < \big(2^{2/3}-1\big)^{3/2}\approx 0.4501964645,
\end{equation}
in which case the perturbation series yields a classical solution $D$.

\subsection{The magnetostatic case}

 When $\rho\equiv 0$ and $D$ vanishes at spatial infinity we have
$D\equiv 0\equiv E$ and the stationary Maxwell--Born--Infeld equations
reduce to the magnetostatic Maxwell--Born(--Infeld) field equations, comprising the stationary
Amp\`ere law
\begin{equation}
        {\nabla}\times H
=
         4 \pi j
\label{eq:AMPERElawAGAIN}
\end{equation}
and the law of the absence of sources of the magnetic induction field $B$,
\begin{equation}
        {\nabla}\cdot B
=
      0
\,.
\label{eq:divBlawAGAIN}
\end{equation}
  The two fields $B$ and $H$ are linked by the aether law of Born \cite{BornA,BornD}, 
\begin{equation}
B 
= 
\frac{H}{ \sqrt{ 1 -\beta^4|H|^2 }} 
\phantom{blablabla}\mathrm{Born's\ B-law}.
\label{eq:BORNlawBofH}
\end{equation}
 While these equations are not directly covered in \cite{CarKieCPAM}, they can be treated
almost verbatim to the treatment  in \cite{CarKieCPAM}.
 The minus sign under the square root is already covered in  \cite{CarKieCPAM}, and otherwise
only the roles of the $\nabla\cdot$ and $\nabla\times$ operators need to be exchanged.

 Thus, inserting our perturbation theory Ansatz $H= H_\Arm + H_\beta$, with 
$H_\beta=\beta^4 H^{(1)}+\beta^8 H^{(2)}+\dots$, into (\ref{eq:BORNlawBofH}) 
and collecting powers of $\beta$, we find that $H^{(k)}$ for $k\in\Nset$ satisfies 
the pair of linear first-order PDE
\begin{eqnarray}
\label{eq:HkCURLlaw}
\nabla\times H^{(k)}&=&0,
\\
\label{eq:HkDIVlaw}
\nabla\cdot H^{(k)}& =&
\nabla\cdot U^{(k)},
\end{eqnarray}
where $U^{(k)}$ is a polynomial in the $H^{(\ell)}$ with $\ell<k$, viz.
\begin{equation}
\label{eq:Udef}
\qquad
U^{(k)} = 
- \sum_{h=1}^k H^{(k-h)}\sum_{j=1}^h |M_j|\!\!
\sum_{|\ell|_{2j} =\atop h-j}\prod_{i=1}^jH^{(\ell_{2i-1})}{\cdot}H^{(\ell_{2i})},
\end{equation}
where again $|\ell|_{2j}$  and $M_j$ have their earlier assigned meaning.
 The pair of linear first-order PDE (\ref{eq:HkCURLlaw}), (\ref{eq:HkDIVlaw}) has
a unique solution in $C^{1,\alpha}_{0}$ given by
\begin{equation}
\label{eq:HkSOLnoD}
H^{(k)} =
{\mathbf Q} U^{(k)},\qquad k\in\Nset
\end{equation}
where $\mathbf{Q}\!: C^{1,\alpha}_{0}\!\to C^{1,\alpha}_{0}$ projects onto the curl-free
subspace of $C^{1,\alpha}_{0}$; see (\ref{eq:HsolEXPLk}).
 The first member of this hierarchy reads explicitly
\begin{equation}
\label{eq:HoneEXPL}
H^{(1)}(x)=
{\textstyle{\frac12}}|H_\Arm|^2H_\Arm
-\nabla\times
\int \frac{{\textstyle{\frac{1}{4\pi}}}\nabla\times({\textstyle{\frac12}}|H_\Arm|^2H_\Arm)(y)}{|x-y|}d^3y,
\end{equation}
so $\beta^4$(\ref{eq:HoneEXPL}) gives the leading correction to the Amp\`ere term $H^{(0)}=H_\Arm$.

 In general, the perturbation series $H= \sum_{k=0}^\infty \beta^{4k} H^{(k)}$, 
with $H^{(0)}=H_\Arm$ given by (\ref{eq:AmpereH}) and $H^{(k)}$ for $k\in\Nset$ 
recursively given by (\ref{eq:HkSOLnoD}), is a \emph{formal} series solution for the
magnetostatic field problem (\ref{eq:AMPERElawAGAIN}), (\ref{eq:divBlawAGAIN})
with prescribed $j$.
 Following nearly verbatim the strategy in \cite{CarKieCPAM} the convergence 
in $C^{1,\alpha}_{0}$ of the formal power series can be proved for 
\begin{equation}
\label{eq:radOFconvAGAIN}
 \beta^2  \|H_\Arm\|  < \big(2^{2/3}-1\big)^{3/2},
\end{equation}
in which case the perturbation series yields a classical solution $H$.

\section{Summary and outlook}

 We have constructed a convergent perturbative power series which solves the stationary 
Maxwell--Born--Infeld equations with regular sources $\rho$ and $j$ in $C^{0,\alpha}_0$
as long as $\beta^2(\|D_\Crm\|+\|H_\Arm\|) <\xi^* = 0.28$ (approximately);
here, $\|\,.\,\|$ is the $C^{1,\alpha}_0$ norm, $D_\Crm$ the Coulomb field and $H_\Arm$
the Amp\`ere field.
 In the absence of either $D_\Crm$ or $H_\Arm$ the corresponding perturbative power series for
the remaining field even converges when $\beta^2\cN< \xi_*\approx 0.45$, 
where $\cN$ is the $C^{1,\alpha}_0$ norm of $H_\Arm$, resp. $D_\Crm$; cf. \cite{CarKieCPAM}.

 While our solution method does not cover the conceptually important case of delta function
sources for which Born conceived of this nonlinear field theory \cite{BornA,BornC,BornInfeldA},
it nevertheless allows one to gain some relevant insight into a burning open question: 
\emph{reliable bounds on the size of Born's aether constant $\beta$.}
 
 In a purely electrostatic calculation Born and Infeld \cite{BornInfeldA} computed 
$\beta =\beta_\Brm \approx 1.236\alpha_\Srm$ (in the units of \cite{KieJSPa,KieJSPb}, 
where $\alpha_\Srm\approx 1/137.036$ is Sommerfeld's fine-structure constant; so,
$\beta_\Brm\approx 0.00902$.
 Subsequently, trying to take the magnetic moment of the electron into account, Born
and Schr\"odinger \cite{BornErwin} came up with a purely magnetostatic (very rough) estimate
of $\beta=\beta_\BSrm\approx 4.74\beta_\Brm$.
 Subsequently Rao \cite{RaoRING} undertook a more detailed electromagnetic study,
following Born's suggestion to treat the electron not as a point but as a charged, rotating 
(one-dimensional) ring. 
 However, all these plausible estimates are not based on an experimentally accessible process, 
involving as they do solely a single ``stationary electron solution'' in an otherwise empty space. 
 Various attempts have been made since to compute effects on atomic spectra induced by the 
Born--Infeld nonlinearity to estimate the range of viable $\beta$ values (see \cite{CarKiePRL} 
and references therein; see also \cite{FranklinGaron} for a recent study), 
but all these methods are based on various as-of-yet uncontrolled approximations
and have produced conflicting results. 

 Our solution method now paves the way for a  systematic study based on controlled approximations.
 For example, if we replace the Dirac delta function of each point charge by a $C^{0,1}$ function
supported on a ball of radius equal to the Compton wavelength of the electron (which sets the 
unit of length in \cite{KieJSPa,KieJSPb}), then $\|D_C\|\approx 3\alpha_\Srm$, so our electrostatic
series will converge for $\beta <_{\!\!\!\!\!{}_\sim} 0.3874/\sqrt{\alpha_S}\approx 4.5348$, which 
range of $\beta$ values safely includes the value proposed by Born. 
 In particular, our perturbation series should converge sufficiently rapidly
so that only a few terms need to be computed; see the appendix.
 Of course, to not compare apples with oranges, the resulting spectral data computed for the 
Schr\"odinger operators obtained with the help of these systematically computed $D=D_\Crm+D_\beta$ 
fields for smeared out charges should be compared with those  computed with the Schr\"odinger 
operator with Coulomb potential for the same ``smeared-out'' charges, not point charges. 

 Similarly, we may also try to estimate the viable range of $\beta$ values for an electron model
with magnetic moment, although comparison with the purely magnetostatic calculation 
by Born--Schr\"odinger will at best be an academic excercise because such a model (presumably)
can't produce the Hydrogen spectrum in the $\beta\to 0$ limit. 
 Instead one should base the assessment on the
computation of a stationary electromagnetic solution with both $\rho$ and $j$ sources
supported on a ``smeared-out'' ring, in the spirit of Rao's study \cite{RaoRING}.
 However, one possible caveat is that the $\beta$ value may get uncomfortably close to the radius of 
convergence of the electromagnetic series, so that more terms may need to be calculated 
than is practically feasible.

 We close the main body of this paper with an outlook on desired generalizations of the  perturbative
solution strategy pursued here for the stationary Maxwell--Born--Infeld field equations in 
three-dimensional Euclidean space.
 The generalization to the stationary model in higher-dimensions, which in recent years has become 
of interest to the high energy theory community (see, e.g., \cite{GibbonsA,LinYang}),
should be quite straightforward (cp. the electrostatic version in \cite{CarKieCPAM}).
 Less straightforward, but very desirable, is the generalization to the genuinely
time-dependent setting; see \cite{SerreA,SerreC,Brenier,Speck} for recent rigorous results,
\cite{Boillat} for an early pioneering study, and \cite{BiBiONE} for a delightful general 
discussion.
 The main problem to be overcome to make the perturbative strategy work in the dynamical setting 
is the identification of a suitable Banach algebra. 
 Lastly, and most importantly from the Born--Infeld perspective \cite{BornA,BornInfeldA},
it is desirable to have the method which can handle delta function sources both in flat space 
(see \cite{BornA,BornInfeldA,Ecker} and \cite{Hoppe,GibbonsA} for some explicit solutions with 
a single, respectively infinitely many point charges) and in curved spacetime 
(see \cite{HoffmannD,HoffmannE,GarciaSalazarPlebanski,Demianski,Breton,ShadiEMBI} 
for explicit solutions with a single point charge).
 The technique presented in this paper can serve to produce seed field-configurations for
growing a solution for delta function sources in a suitable limiting process. 
 Progress in this direction will be reported elsewhere.

\smallskip
\noindent
\textbf{Acknowledgment:} Work supported by the NSF through Grant DMS-0807705. 
Any opinions expressed in this paper are those of the author and do not necessarily
reflect those of the NSF. 
 I thank Shadi Tahvildar-Zadeh and Holly Carley for their comments. 
 Special thanks are due to Doron Zeilberger and Shalosh B. Ekhad 
\cite{Ekhad} for their wizzardry help with the recursion relation for the coefficients $R_{k}$.
\newpage

\appendix
\section*{Appendix\footnote{\textbf{Disclaimer}: 
Everything in this appendix could be made rigorous, but this here is only 
semi-$\phantom{\qquad 1Disclaimer:}$rigorous. The whole process took Ekhad 200.542 seconds.}}

 Practically only a small number of terms of the perturbation series for $D$ and $H$ 
can be computed.
 Here we supply some details about the recursion coefficients $R_k$
which are needed in the error estimates.
 Given the input value $R_0=1$, the author computed $R_1,R_2,R_3$ by hand from the recursion
formula given in the main text, then used MAPLE to compute 14 terms using the iteration of the fixed 
point map given by the implicit definition of $G(\xi)$ in a direct, 
unpolished manner (after which MAPLE reached its capacity). 
 The more sophisticatedly iterating Shalosh B. Ekhad knows 303 terms of the sequence 
$\{R_{k}\}_{k\in\Nset\cup\{0\}}$; here are the first 30 of them:

\vskip-.7truecm
\begin{eqnarray}
R_{0}\ &=&1\\
R_{1}\ &=&\frac{3}{2}\\
R_{2}\ &=&\frac{65}{8}\\
R_{3}\ &=&\frac{943}{16}\\
R_{4}\ &=&\frac{62689}{128}\\
R_{5}\ &=&\frac{1128197}{256}\\
R_{6}\ &=&\frac{42790845}{1024}\\
R_{7}\ &=&\frac{842157399}{2048}\\
R_{8}\ &=&\frac{136312116961}{32768}\\
R_{9}\ &=&\frac{2817640708457}{65536}\\
R_{10}&=&\frac{118490151386655}{262144}\\
R_{11}&=&\frac{2526390089218393}{524288}\\
R_{12}&=&\frac{217977129447815405}{4194304}
\end{eqnarray}

\begin{eqnarray}
R_{13}&=&\frac{4748017126294329161}{8388608}\\
R_{14}&=&\frac{208584441836961908949}{33554432}\\
R_{15}&=&\frac{4614991020517094410279}{67108864}\\
R_{16}&=&\frac{1644116252728526666074977}{2147483648}\\
R_{17}&=&\frac{36812969231234813601419473}{4294967296}\\
R_{18}&=&\frac{1656740336870818323274233515}{17179869184}\\
R_{19}&=&\frac{37445969415289365495538129125}{34359738368}\\
R_{20}&=&\frac{3398982473269915594232889691503}{274877906944}\\
R_{21}&=&\frac{77410530113072758320538102052283}{549755813888}\\
R_{22}&=&\frac{3537571318060518004220386126288923}{2199023255552}\\
R_{23}&=&\frac{81073698522885193141789999978889377}{4398046511104}\\
R_{24}&=&\frac{14905122955618940253574385037312323437}{70368744177664}\\
R_{25}&=&\frac{343396823100629008332240991968973221221}{140737488355328}\\
R_{26}&=&\frac{15859792436328056179243840618808531803115}{562949953421312}\\
R_{27}&=&\frac{367032135637139188746851720898633881475805}{1125899906842624}\\
R_{28}&=&\frac{34043872901750574940303463222092806831775365}{9007199254740992}\\
R_{29}&=&\frac{790900804799227563339255535454710815651794705}{18014398509481984}.
\end{eqnarray}

 For many practical purposes the above list should be fully sufficient.
 However, should more terms be needed it is more efficient to switch to an 
asymptotic representation. 
 Unfortunately, the recursion formula found in this paper is not practically useful for
this purpose, for it quickly grows out of control because it involves \emph{all} previous
$R_\ell$ values ($\ell<k$) to compute the $k$-th one.
 And the fix point iteration produces the $R_k$ without yielding control of their asymptotics.
 Fortunately, we know from general theorems that the associated sequence of integers
$\{S_{k+1}:= 2^{2k}R_k\}_{k\in\Nset\cup\{0\}}$ satisfies a linear recurrence of fixed finite 
order with coefficients which are fixed polynomials in $n=k+1\in\Nset$.
 The beginning of this integer sequence reads $\{1,6,130,3772,125378,...\}$. 
 After a few minutes of interaction with Doron Zeilberger, Shalosh B. Ekhad found:
given $\{S_1,...,S_9\}$, the ensuing $S_n$ values are obtained as
\begin{equation}
\label{eq:LINEARrecursion}
S_{n+1} = {\sum}_{h=0}^8 P_h(n)S_{n-h},
\end{equation}
where each $P_h$ is a polynomial of degree 9 in $n$, with integer coefficients. 
 The coefficients of these polynomials are insanely large, with more than 100 digits, so that
it would be pointless to list them here; they are available from Dr.Z.'s computer.

 The asymptotic behavior of $\{S_{k+1}\}_{k\in\Nset\cup\{0\}}$ can now be extracted from 
the linear recurrence.
 The leading term in the asymptotics is determined by the degree-$8$ polynomial in the 
complex variable $z$ formed by collecting the coefficients of highest power (i.e. $n^9$)
of the polynomials $P_h(n)$ in (\ref{eq:LINEARrecursion}), each multiplied by the associated 
$h$-th power of $z$, thus (symbolically)
\begin{equation}
\label{eq:LEADpolyINz}
\widetilde{P}(z) =  {\sum}_{h=0}^8 {\textstyle{\frac{1}{9!}}}P_h^{(9)}(n) z^h,
\end{equation}
where $P_h^{(9)}(n)$ is the $9$-th derivative of $P_h(n)$, with $n$ formally treated as real
variable.
 As $n\to\infty$, the ratio $S_{n+1}/S_n$ converges to the largest root of (\ref{eq:LEADpolyINz}),
which Ekhad easily finds as $\approx 48.9391511596$ (truncated after 10 digits).
 Dividing by 4 yields 
\begin{equation}
\label{eq:limitRATIO}
 \lim_{k\to\infty} {R_{k+1}}/{R_k} = q \approx12.23478778990.
\end{equation}
 As many digits of the ratio $q$ are practically reached only for $k\gg 30$;
however, empirically, ${R_{k+1}}/{R_k}$ is monotonic increasing, and so
\begin{equation}
 {R_k} < q^k.
\end{equation}
 This suffices to obtain the relevant error estimates for the truncation of the series.

 Incidentally, by the ratio criterion, (\ref{eq:limitRATIO}) yields the radius of convergence of 
$G(\xi)$ as $\xi^* = 1/\sqrt{q} \approx 0.285891853$, reproducing what we found earlier quasi-algebraically.
\newpage

\small
\bibliographystyle{plain}


\end{document}